\documentclass{aa}

\usepackage{harvard,times}

\newcommand{\lesssim}{\mathrel{\hbox{\rlap{\hbox{\lower4pt\hbox{$\sim$}}}\hbox{$<$}}}}
\newcommand{\gtrsim}{\mathrel{\hbox{\rlap{\hbox{\lower4pt\hbox{$\sim$}}}\hbox{$>$}}}}

\begin{document}

%\msnr{A\&A/1999/9120 - in press}

\thesaurus{2(02.01.2; 02.08.1; 02.20.1; 08.16.5; 10.05.1;
11.05.2)}

\title{A Note on Hydrodynamic Viscosity and Selfgravitation\\
in Accretion Disks}

\author{Wolfgang J.\ Duschl \inst{1,3,}\thanks{wjd@ita.uni-heidelberg.de}
\and Peter A.\ Strittmatter \inst{2,3,}\thanks{pstrittmatter@as.arizona.edu}
\and Peter L.\ Biermann \inst{3,}\thanks{p165bie@mpifr-bonn.mpg.de}}

\institute{Institut f\"ur Theoretische Astrophysik,
Tiergartenstr.\ 15, 69121 Heidelberg, Germany \and Steward
Observatory, The University of Arizona, Tucson, AZ 85721, USA
\and Max-Planck-Institut f\"ur Radioastronomie, Auf dem H\"ugel
69, 53121 Bonn, Germany}

\offprints{W.J.\ Duschl, Institut f\"ur Theoretische Astrophysik,
Tiergartenstr.\ 15, 69121 Heidelberg, Germany}

\mail{W.J.\ Duschl, Institut f\"ur Theoretische Astrophysik,
Tiergartenstr.\ 15, 69121 Heidelberg, Germany;\\
wjd@ita.uni-heidelberg.de}

\date{Received / Accepted}

\maketitle

\begin{abstract}
We propose a generalized accretion disk viscosity prescription
based on hydrodynamically driven turbulence at the critical
effective Reynolds number. This approach is consistent with
recent re-analysis by \citeasnoun{RZa99} of experimental results
on turbulent Couette-Taylor flows. This new $\beta$-viscosity
formulation is applied to both selfgravitating and
non-selfgravitating disks and is shown to yield the standard
$\alpha$-disk prescription in the case of shock dissipation
limited, non-selfgravitating disks. A specific case of fully
selfgravitating $\beta$-disks is analyzed. We suggest that such
disks may explain the observed spectra of protoplanetary disks and
yield a natural explanation for the radial motions inferred from
the observed metallicity gradients in disk galaxies. The
$\beta$-mechanism may also account for the rapid mass transport
required to power ultra luminous infrared galaxies.


\end{abstract}

\section{Introduction\label{introduction}}

One of the major shortcomings of the current theoretical
descriptions of accretion disks is lack of detailed knowledge
about the underlying physics of viscosity in the disk. This
problem is significant because almost all detailed modelling of
the structure and evolution of accretion disks depends on the
value of the viscosity and its dependence on the physical
parameters. There is general agreement that molecular viscosity
$\nu_{\rm mol}$ is totally inadequate and that some kind of
turbulent viscosity is required. Moreover, the Reynolds number in
the disk flow is extremely high in any astrophysical context and
this in itself is likely to lead to strong turbulence regardless
of the details of the actual instability involved.

However, there is far less certainty about how to prescribe such
a turbulent viscosity in the absence of a proper physical theory
of turbulence. Most investigators adopt the so-called
$\alpha$-{\it ansatz\/} introduced by \citeasnoun{Sha72} and
\citeasnoun{SSu73} that gives the viscosity ($\nu$) as the
product of the pressure scale height in the disk ($h$), the
velocity of sound ($c_{{\rm s}}$), and a parameter $\alpha$ that
contains all the unknown physics:

\begin{equation}
\label{viscosity}
\nu = \alpha h c_{\rm s} .
\end{equation}
One interprets this as some kind of isotropic turbulent viscosity
$\nu = \nu_{{\rm t}} = l_{{\rm t}} v_{{\rm t}}$ where $l_{{\rm
t}}$ is an ({\it a priori\/} unknown) length scale and $v_{{\rm
t}}$ an ({\it a priori\/} unknown) characteristic velocity of the
turbulence. One may then write $\alpha = (v_{{\rm t}}/c_{{\rm
s}})\cdot(l_{{\rm t}}/h)$. On general physical grounds neither
term in parentheses can exceed unity so that $\alpha \le 1$. If
initially $v_{{\rm t}} > c_{{\rm s}}$, shock waves would result
in strong damping and hence a return to a subsonic turbulent
velocity. The condition $l_{{\rm t}} > h$ would require
anisotropic turbulence since the vertical length scales are
limited by the disk's thickness, which is comparable to $h$.

While it is always, in a trivial way, possible to calculate a
value $\alpha$, a parameterization of this sort for $\nu$ is only
useful if the proportionality parameter, $\alpha$, is
(approximately) constant. One can expect this to happen only if
the scaling quantities are chosen in a physically appropriate
manner. Models for the structure and evolution of accretion disks
in close binary systems (e.g., dwarf novae and symbiotic stars)
show that Shakura \& Sunyaev's parameterization with a constant
$\alpha$ leads to results that reproduce the overall observed
behaviour of the disks quite well. Time dependent model
calculations of the outbursts of dwarf novae
\citeaffixed{MMH84}{e.g.,} and X-ray transients
\citeaffixed{Can96}{e.g.,} demonstrate that, over a wide range of
physical states of a disk in different phases of its evolution,
the derived values of the viscosity parameter $\alpha$ do not
vary by more than approximately an order of magnitude and are not
too different from unity. As a result of this success, the
$\alpha$-ansatz is now used in essentially all accretion disk
models.

It is noteworthy that, despite this success, the $\alpha$-ansatz
retains no information about the mechanism generating the
turbulence but only about physical limits to its efficiency in a
disk. In fact we would expect any high Reynolds number
astrophysical shear flow to exhibit some kind of turbulent
viscosity regardless of whether or not it happens to be in a
disk. We therefore conclude that a more general prescription
underlies the $\alpha$-ansatz for accretion disks.

In recent years, \citeasnoun{BHa91} and their collaborators
\citeaffixed{HGB95}{e.g.,} have shown that for
non-selfgravitating magnetic accretion disks, an instability
exists that can give rise to turbulence with the required formal
dependence and---if only marginally----the required amount. We
also note that in substantial regions of proto-stellar and
proto-planetary disks, the charge density is unlikely to be high
enough to sustain a significant magnetically mediated viscosity,
although this phenomenon may be relevant elsewhere.

Whether a purely hydrodynamic turbulence can sustain the
viscosity in the angular momentum profile of an accretion disk
and can result in an angular momentum transport towards regions
with larger specific angular momentum is still a matter of
debate. \citeasnoun{BHS96}, for instance, argue against it, based
on numerical experiments, albeit for a rather low effective
Reynolds number. \citeasnoun{Dub92} and \citeasnoun{KYo97}, among
others, argue in favor of it, mainly based on analytical
considerations. Experiments dating back to the 1930s on the
Couette-Taylor flow between co-axial rotating fluids
\cite[b]{Wen33,Tay36a} show clearly the existence of a purely
hydrodynamic instability. While the flow is essentially
incompressible, turbulence is generated above a critical Reynolds
number, independent of the radial profile of angular
momentum\footnote{One of the cases investigated by Wendt indeed
has a rotation law which approximates closely the profile in a
Keplerian disk.}. A modern review has been given by
\citeasnoun{DPS85}. Most recently \citeasnoun{RZa99} (hereafter
RZ) have undertaken a reanalysis of Taylor's experimental results
and, for high Reynolds number flow, have interpreted them in
terms of a turbulent viscosity (see also Sect.\
\ref{solutioneins}).

In this contribution, we adopt the view that hydrodynamically
driven turbulence can sustain the viscosity in accretion disks.
We suggest, in Sect.\ \ref{solutioneins}, a viscosity
prescription, the $\beta$-ansatz, that represents the maximum
attributable to hydrodynamic turbulence. We show that in the
limit of low mass, thin disks, hydrodynamic turbulence will
result in the Shakura-Sunyaev prescription. We then discuss the
implications of the proposed formulation for the structure and
evolution of selfgravitating disks, noting that even for these
disks, the viscosity prescription differs from the
$\alpha$-ansatz and hence removes a difficulty first noted by
\citeasnoun{Pac78}. Finally, we discuss protostellar, galactic
and galactic center disks as examples where the $\beta$-ansatz
may be relevant.

\section{Prescription for Turbulent Viscosity\label{solutioneins}}

\subsection{Reynolds viscosity as the general case}

As noted in Sect.\ \ref{introduction}\ the need for some kind of
turbulent viscosity in accretion disks is generally recognized,
as is the very high Reynolds number of the flow in the absence of
such a viscosity. Here, we wish to investigate in particular the
case of accretion disks where the magnetic fields do not play an
important role. In these circumstances, it seems reasonable to
assume that the turbulence is driven by the velocity field in the
disk, which itself has characteristic length and velocity scales
$s$ (the radius of the orbit) and $v_\phi$ (the azimuthal
velocity), respectively.

As has been pointed out, for example, by \citeasnoun{LBP74} and
\citeasnoun{TSE77}, the high corresponding Rey\-nolds number $\Re
= s v_\phi / \nu$ should lead to the generation of turbulence and
hence to a steady enhancement in the effective viscosity. This
will continue until the Reynolds number has been reduced to
approximately its critical value $\Re_{\rm crit}$. Typical values
for $\Re_{\rm crit}$ in laboratory flows are of the order of
$\sim 10^2-10^3$. This limiting Reynolds viscosity can, in this
case, be as high as

\begin{equation}\label{betaansatz}
\nu = \nu_{\rm R} = \beta s v_\phi
\end{equation}
where $\beta$ is a constant satisfying

\begin{equation}\label{betalimit}
\beta \lesssim \frac{1}{\Re_{\rm crit}} \sim 10^{-3} - 10^{-2}.
\end{equation}
In terms of previously introduced quantities we may write $v_{\rm
t} \sim \beta_1 v_\phi$ and $l_{\rm t} \sim \beta_2 s$ so that
$\beta \sim \beta_1 \beta_2 \sim 10^{-2 \dots -3}$.

In support of this choice of $s$ as the natural length scale we
note that it is the only length scale which is relevant for
angular momentum transport and which contains information about
the driving agent for the turbulence---namely the rotation field;
likewise, the orbital velocity $v_{\phi}$ is the only velocity
scale containing such information.

This approach receives further support from the reanalysis by RZ
of the \citeasnoun{Wen33} and \citeasnoun[b]{Tay36a} experiments
on turbulent viscosity generated in the flow between coaxial
rotating cylinders. We note, however, that it is difficult to
make precise comparisons between accretion disk and rotating
cylinders in view of quite different constraints on the fluid
flow.

Using a definition of $\Re \sim R \Delta\Omega \Delta R / \nu$
appropriate to the experimental situation (here $R$ is the
average cylinder radius and $\Delta\Omega$ and $\Delta R$ are the
relative angular velocity and gap size between the cylinders), RZ
derive expressions for $\Re_{\rm crit}$ as a function of relative
gap size $\Delta R / R$. For small gap size they find $\Re_{\rm
crit} \lesssim 2000$ independent of gap size. For large relative
gap size they find  that $\Re_{\rm crit} = \Re_{\rm grad} (\Delta
R/R)^2$ where their {\it gradient Reynolds number\/} $\Re_{\rm
grad} \sim r^3 (d\Omega/d R) \sim 10^6$ is essentially constant.
Thus for small gaps the experimental data yield essentially the
same value of $\Re_{\rm crit}$ as we have adopted in Eq.\
(\ref{betalimit}). For large gap sizes, the constancy of
$\Re_{\rm grad}$ leads to essentially the same {\it functional\/}
form as in Eq.\ (\ref{betaansatz}) but with a significantly
smaller value of $\beta$. RZ arrive at similar conclusions for
the two regimes of gap sizes from an analysis of the torques
exerted on the cylinders in cases where the flow is turbulent. We
believe that the small gap limit is the more relevant to the
accretion disk case, for which the speed of rotation is
constrained at each radius by the gravitational field. The flux
of angular momentum at each radius is thus determined by the
imposed orbital velocity field. By contrast, the experimental
configuration constrains the flow velocity only at the inner and
outer radius with the flow in the gap region able to take up a
velocity profile determined by viscosity and in which the angular
momentum flux is independent of radius. In our opinion the
experimental results provide strong support for a turbulent
viscosity generated by hydrodynamically driven turbulence. While
we recognize that more work is required on the question, we also
believe that the small gap results provide significant support
for the viscosity prescription given in Eqs.\ (\ref{betaansatz})
and (\ref{betalimit}).

In the following we will refer to Eq.\ \ref{betaansatz} as the
$\beta$-ansatz and to the disk structure arising from this
viscosity prescription as $\beta$-disks. We suggest that the
$\beta$-ansatz is the most appropriate initial formulation for
accretion disks since it is directly connected to the driving
mechanism. It establishes the maximum value of the viscosity that
can arise from hydrodynamically driven turbulence.

The actual viscosity may, however, be limited to lower values by
such phenomena as shock dissipation of turbulent energy if the
implied turbulent velocities exceed the local sound speed. As we
show in Sect.\ \ref{alphalimit}, this yields the $\alpha$-ansatz
in conditions relevant to these non-selfgravitating accretion
disks. However, it leads to a different prescription in shock
limited selfgravitating disks. In the Couette-Taylor case, all
velocities were subsonic (the flow was essentially
incompressible) so that no additional constraints applied. This
would also be the case in astrophysical disks in which $\beta_1 <
c_{\rm s}/v_\phi$.

\subsection{$\alpha$-viscosity as the limiting case for shock
dissipation limited low mass accretion disks.\label{alphalimit}}

If the accretion disk is such that the local sound speed is less
than the turbulent velocities implied by the $\beta$-ansatz,
i.e., $\beta_1 > c_{\rm s}/v_\phi$, we may rewrite Eq.\
\ref{betaansatz} as

\begin{equation}
\nu = v_{\rm t} l_{\rm t} \sim \Delta v_\phi \Delta s
\end{equation}
where $\Delta v_\phi$ and $\Delta s$ are the maximum
representative velocity and length scales allowed by local
conditions. Furthermore, we may write

\begin{equation}\label{deltavphi}
\Delta v_\phi = \frac{\partial v_\phi}{\partial s} \Delta s \sim
\frac{v_\phi}{s} \Delta s
\end{equation}
so that a restriction on $\Delta v_\phi$ implies a constraint
also on $\Delta s$ and vice-versa.

If we consider turbulent elements in a {\it smoo\-thed out\/}
background gas with sound speed $c_{\rm s}$ we may impose the
limit that the turbulent velocity will approach but may not
exceed $c_{\rm s}$. Thus Eq.\ \ref{deltavphi} gives

\begin{equation}
\frac{v_\phi}{s} \Delta s \sim \Delta v_\phi = \zeta c_{\rm s}
\end{equation}
or

\begin{equation}\label{deltasgeneral}
\Delta s \sim \frac{s}{v_\phi} \Delta v_\phi = \zeta
\frac{s}{v_\phi} c_{\rm s}
\end{equation}
with $\zeta$ a quantity smaller than but of order unity. This
estimate of $\Delta s$ may be interpreted as the distance a
hydrodynamically driven turbulent element can travel before
loosing its identity due to shock dissipation.

In a standard geometrically thin, non-selfgra\-vi\-ta\-ting
accretion disk (i.e., a Shakura-Sunyaev, or $\alpha$-disk)
hydrostatic equilibrium in the vertical direction implies

\begin{equation}
\label{hydrostaticnsg}
\frac{h}{s} = \frac{c_{\rm s}}{v_\phi}
\end{equation}
Using this in Eq.\ \ref{deltasgeneral}, we find for
Shakura-Sunyaev disks

\begin{equation}\label{deltasnsg}
\Delta s \sim \zeta h
\end{equation}
and hence that

\begin{equation}\label{nunsg}
\nu \sim \alpha h c_{\rm s}
\end{equation}
with $\alpha = \zeta^2$ again not too much smaller than unity.
This derivation of the Shakura-Sunyaev scaling, starting from the
assumption of a Reynolds driven turbulence, depends on the disk
mass being negligible, i.e., a vertical hydrostatic equilibrium of
the form of Eq.\ \ref{hydrostaticnsg} has to apply. For
selfgravitating disks, Eq.\ \ref{hydrostaticnsg} no longer
applies, and thus the functional form of $\nu$ will differ from
that of Eq.\ \ref{nunsg}. Note that the upper bound to $\Delta s$
{\it implies\/} approximately isotropic turbulence. This is the
standard $\alpha$-ansatz but derived from considerations of
rotationally generated turbulence.

It is worthwhile noting that this derivation of the
Shakura-Sunyaev prescription not only yields its functional form,
$\nu \propto h c_{\rm s}$, but also the order of magnitude for
the scaling parameter, $\alpha \sim \zeta^2$, where $\alpha$ is
close to but less than unity. This value is consistent with
values derived by comparing $\alpha$-disk models with
observations of disks, for instance in dwarf novae \cite{CSW88}.

From the above, it is clear that the viscosity in accretion disks
depends not only on the generation of hydrodynamic turbulence but
also on the limitation arising from the requirement that the
turbulence be subsonic. It also depends on whether or not the
disk is selfgravitating. In the following Sect., we use he same
logic to investigate the viscosity prescription in
selfgravitating accretion disks, in which turbulence is limited
by shock dissipation.

\section{Viscosity in Thin Selfgravitating Accretion
Disks\label{sectstructure}}

\subsection{Conditions for Selfgravity in Accretion
Disks\label{sectsg}}

In the following we assume that the accretion disks are
geometrically thin in the vertical direction, symmetric in the
azimuthal direction, and stationary. We approximate the vertical
structure by a one zone model. Then a disk model is specified by
the central mass $M_*$, the radial distributions of surface
density $\Sigma(s)$, central plane temperature $T_{{\rm c}}(s)$,
and effective temperature $T_{{\rm eff}}(s)$ or the radial mass
flow rate $\dot M$. The relevant material functions are the
equation of state, the opacity and the viscosity prescription.

One can estimate the importance of selfgravity by comparing the
respective contributions to the local gravitational accelerations
in the vertical and radial directions.

The vertical gravitational acceleration at the disk surface is $2
\pi G \Sigma$ and $G M_* h / s^3$, for the selfgravitating and
the purely Keplerian case, respectively. Selfgravitation is thus
dominant in the vertical direction when

\begin{equation}
\label{eqcondvertsg}
\frac{M_{\rm d}}{M_*} \sim \frac{\pi s^2 \Sigma}{M_*} > \frac{1}{2}
\frac{h}{s},
\end{equation}
where $M_{\rm d}(s)$ is the mass enclosed in the disk within a
radius $s$ and is given approximately by $M_{\rm d} \sim \pi s^2
\Sigma$. Typical numbers for $h/s$ are in the range $10^{-2 \dots
-1}$.

Similar considerations lead to the condition

\begin{equation}
\label{eqcondradsg}
M_{\rm d} > M_*
\end{equation}
for selfgravitation to dominate in the radial direction. Thus, we
can define three regimes as follows:
\begin{itemize}
\item Non-selfgravitating (NSG) disks in which $M_{\rm d}(s)$
$\le (1/2) (h/s) M_*$ (i.e., the classical Shakura-Sunyaev disks)
\item Keplerian selfgravitating (KSG) disks in which selfgravity is
significant only in the vertical direction and which satisfy the
constraint $(1/2) (h/s) $ $M_* \le M_{\rm d}(s) \le M_*$
\item Fully selfgravitating (FSG) disks which satisfy $M_* \le
M_{\rm d}(s)$
\end{itemize}

Because $M_{\rm d}(s)$ is a monotonically increasing function of
$s$, all three regimes will arise in sufficiently massive, thin
($h/s \ll 1)$ disks.

\subsection{Selfgravitating
Disks\label{inconsistency}\label{structure}}

In this Sect., we will review the structure of selfgravitating
(SG) accretion disks within the framework of the assumptions
introduced above. Compared to the standard NSG models, both the
KSG and FSG disks require modification of the equation of
hydrostatic support in the direction perpendicular to the disk.
Thus, while in the standard model the local vertical pressure
gradient is balanced by the $z$ component of the gravitational
force due to the central object, in the SG case we have balance
between two local forces, namely the pressure force and the
gravitational force due to the disk's local mass. In the KSG
case, in the radial direction centrifugal forces are still
balanced by gravity from a central mass ({\it Keplerian\/}
approximation), while in the fully selfgravitating case we have
to solve Poisson's equation for the rotation law in the disk.

For an SG disk, hydrostatic equilibrium in the vertical direction
yields

\begin{equation}  \label{hydrostaticequilibrium}
P = \pi G \Sigma^2
\end{equation}
\cite{Pac78}, where $P$ is the pressure in the central plane ($z
= 0$), $\Sigma$ is the surface mass density integrated in the $z$
direction, and $G$ is the gravitational constant.

Since details of the thermodynamics in the $z$ direction are of
no particular relevance to our argument, we shall assume the disk
to be isothermal in the vertical direction.

Integrating the equation of conservation of angular momentum gives

\begin{equation}  \label{angularmomentum}
\nu \Sigma = - \frac{\dot M}{2 \pi s^3 \omega^\prime}(s^2 \omega
- \xi)
\end{equation}
with the radial mass flow rate\footnote{We choose the convention
of radial mass flow rate $\dot M$ and radial velocity $v_s$ {\it
positive\/} for inward motion.} $\dot M$, the rotational
frequency $\omega$, its radial derivative $\omega^\prime$, and a
quantity $\xi$ allowing for the integration constant or,
equivalently, for the inner boundary condition. For a detailed
discussion of $\xi$ see, e.g., \citeasnoun{DTs91},
\citeasnoun{PNa95} and \citeasnoun{DBi96}. For simplicity, we set
the boundary condition $\xi = 0$ in the subsequent discussion.
This does not alter the essence of our argument, and only changes
details close to the disk's inner radial boundary, since the
product $s^2\omega$ increases with $s$. In fact Eq.\
\ref{angularmomentum} applies in the general case (i.e., NSG and
SG); for Keplerian disks (NSG and KSG), we may write
$\omega^\prime = - 3 \omega / 2 s$.

Finally, we have for the sound velocity

\begin{equation}  \label{soundvelocity}
c_{{\rm s}}^2 = P \left/ \left( \frac{\Sigma}{2 h} \right) \right. .
\end{equation}
Eqs.\ \ref{hydrostaticequilibrium} and \ref{soundvelocity} give

\begin{equation}  \label{pgsh}
2 \pi G \Sigma h = 4 \pi G \overline{\rho} h^2 = c_{{\rm s}}^2
\end{equation}
where $\overline{\rho} = \Sigma / 2 h$ is a vertically averaged
mass density.

On the other hand, the Jeans condition for fragmentation in the
disk into condensations of radius $R$ is

\begin{equation}  \label{jeans}
\frac{4\pi}{3} q G \overline{\rho} R^2 > c_{{\rm s}}^2
\end{equation}
\citeaffixed{Mes65}{see} where $q$ is factor of order unity.

Thus, a selfgravitating disk is on the verge of fragmenting into
condensations of radius $R \sim h$ unless these are destroyed by
shear motion associated with the Keplerian velocity field. Thus
\citeasnoun{Pac78} and later \citeasnoun{KWP79} and
\citeasnoun{LPr87} proposed that the viscosity prescription was
directly coupled to the above gravitational stability criterion.

To solve for the dynamic and thermal structure of SG disks, a
viscosity prescription has to be specified. As in the NSG case it
is possible but not necessary that viscosity is limited by shock
dissipation. In the absence of such dissipation, we would, as
before, expect the $\beta$-viscosity to apply. It is instructive,
however, to follow the logic of Sect. \ref{alphalimit} in he case
of SG disks in which turbulence is limited by shock dissipation.

\subsection{Viscosity in shock dissipation limited\\
selfgravitating accretion disks \label{shocklimsg}}

For a SG disk (whether Keplerian or not) Eq.\
\ref{hydrostaticnsg} is no longer valid so the analysis of Eqs.\
\ref{deltasnsg} and \ref{nunsg} no longer applies. In physical
terms, the scale height in the disk no longer reflects global
properties of the disk (mass of and distance to the central star)
but is set by local conditions.

For the selfgravitating case we have approximately

\begin{equation}
v_\phi^2 \sim \frac{G ( M_* + M_{\rm d})}{s}
\end{equation}
where conditions $M_* \gg M_{\rm d}$, and $M_* \ll M_{\rm d}$.
distinguish between the Keplerian and the FSG cases,
respectively, with the KSG disks as an intermediate case.

From Eq.\ \ref{deltasgeneral}

\begin{equation}\label{deltassg}
\Delta s \sim \zeta \frac{s c_{\rm s}}{v_\phi} \sim \zeta \left(
\frac{2 \pi h \Sigma s^3}{M_* + M_{\rm d}} \right)^{1/2} \sim
\zeta \left( \frac{2 h \Sigma s}{\Sigma_* + \Sigma} \right)^{1/2}
\end{equation}
where $\Sigma_*$ is defined by $M_* = \pi s^2 \Sigma_*$.

At the transition to the NSG regime, Eqs.\ \ref{eqcondvertsg} and
\ref{deltassg} and the condition $M_* \gg M_{\rm d}$ give as
before

\begin{equation}
\Delta s \sim \zeta h
\end{equation}
and hence a smooth transition to the $\alpha$-ansatz.

For the FSG regime, Eq.\ \ref{deltassg} and the condition $M_{\rm
d} \gg M_*$ give the simple asymptotic form

\begin{equation}
\Delta s \sim \zeta ( 2 h s )^{1/2}
\end{equation}
and hence a viscosity of the form

\begin{equation}
\nu \sim \gamma ( h s )^{1/2} c_{\rm s}
\end{equation}
where $\gamma$ is a factor of order unity.

The situation is more complex for intermediate values of $M_{\rm
d}/M_*$. The derived viscosity differs in all SG cases from the
standard $\alpha$-ansatz, but approaches that form as $M_{\rm
d}/M_* \rightarrow 0$. Thus when hydrodynamically induced
turbulence is limited by shock dissipation, the resultant
viscosity reflects local conditions and takes the standard
Shakura-Sunyaev form only when the disk mass is negligible. We
show below that this new prescription removes a problem
previously noted by \citeasnoun{Pac78} and others, with the
structure of KSG disks with $\alpha$-viscosity.

\subsection{Structure of SG Disks with Shock Limited Viscosity}

It follows from Eqs. \ref{hydrostaticequilibrium},
\ref{angularmomentum} and \ref{soundvelocity}, with $\xi = 0$,
that

\begin{equation}
c_{\rm s}^2 = - \frac{G h \dot{M}}{2 \nu} \left( \frac{ {\rm d}
\ln s}{ {\rm d} \ln \omega} \right). \label{csquadrat}
\end{equation}
If one adopts the standard $\alpha$-prescription, this yields

\begin{equation}
c_{\rm s}^3 = - \frac{G \dot{M}}{2 \alpha} \left( \frac{ {\rm d}
\ln s}{ {\rm d} \ln \omega} \right).
\end{equation}

For a KSG disk, this in turn yields

\begin{equation}
c_{\rm s}^2 = \left( \frac{2 G \dot{M}}{3 \alpha} \right)^{2/3} =
\frac{ k T_{\rm c}}{m_{\rm H}},
\end{equation}
or
\begin{equation}
\label{eqtconst} T_{\rm c} = 2.42\,{\rm K} \left( \frac{1}{\alpha}
\frac{\dot{M}}{10^{-6}\,{\rm M}_\odot/{\rm yr}}\right)^{2/3}
\end{equation}
A similar result arises for the FSG case, albeit with a different
numerical factor resulting from the solution of Poisson's
equation.

Thus, for a SG disk, the $\alpha $-ansatz leads to the
requirement of a constant temperature for all radii $s$ (or, if
$\xi \not = 0$ in Eq.\ \ref{angularmomentum}, the temperature is
prescribed as a function of $s$), independent of thermodynamics.
While the \underline{exact} constancy of the temperature may very
well be an artefact of our simplified one-zone approximation for
the vertical structure, there is no reason to expect that proper
vertical integration of the structure will change this
fundamentally.

The $\alpha$ ansatz for a SG disk also requires that the disk
structure satisfy

\begin{equation}  \label{hsigma}
h \Sigma = \frac{c_{{\rm s}}^2}{2 \pi G} = \frac{\dot
M^{2/3}}{\left( 3 \alpha \right)^{2/3} \left( 2 \pi G
\right)^{1/3} }.
\end{equation}
In a standard NSG accretion disk the temperature is a free
parameter which is determined by the energy released by the
inward flow of the disk gas ($\dot M$), by the local viscosity,
and by the respective relevant cooling mechanisms. The viscosity
depends on $T_{{\rm c}}$ via Eq.\ (\ref {viscosity}) and on the
equation of hydrostatic support in the direction normal to the
disk (Eq.\ (\ref{hydrostaticnsg}), which in the non-SG case
replaces Eq.\ (\ref {hydrostaticequilibrium})).

In the SG case, it is the surface density $\Sigma$ (and hence
$h$) which must adjust in order to radiate the energy deposited
by viscous dissipation and provided by the inward flowing
material. While detailed solutions are beyond the scope of this
paper, they clearly exist formally. On the other hand, the normal
{\it thermostat\/} mechanism does not operate, at least in the
steady state. Indeed in certain circumstances, the condition of
constant mid-plane temperature appears to be inconsistent with
the basic thermodynamic requirement that the average gas
temperature in the disk exceed that of the black body temperature
required to radiate away the energy dissipated by viscous
stresses (see Appendix). It is therefore doubtful whether a
physically plausible and stable quasi-steady state solution
exists.

On the other hand, if one adopts the alternative prescription for
shock limited viscosity proposed in Sect. \ref{shocklimsg}, the
above problem with constant or prescribed mid-plane temperature
disappears, the temperature once again depends on $h$, and the
normal thermostat can operate. While this does not prove the
validity of the shock limited viscosity prescription given in
Sect.\ \ref{shocklimsg}, it is certainly an interesting
consequence.

\section{Selfgravitating $\beta$-Disks}

\subsection{General Observations}

A general analysis of SG disks is complex and beyond the scope of
this paper. In this Sect., we examine the structure of
$\beta$-disks, in which the turbulence is subsonic at all radii.
Before doing so, we make the following general observations.

First, with the $\beta$-viscosity prescription, Eqs.\
\ref{angularmomentum} and \ref{csquadrat} give

\begin{equation}  \label{sigmabeta}
\Sigma = - \frac{\dot{M} \omega }{2\pi\nu s \omega^\prime} = -
\frac{\dot M}{\beta s^3 \omega^\prime}
\end{equation}
and

\begin{equation}
c_{{\rm s}}^2 = - \frac{G h \dot{M}}{2 \nu} \frac{\omega}{s
\omega^\prime} = - \frac{G h \dot M}{2 \beta s^3 \omega^\prime}
\label{csquadratsg}
\end{equation}
Thus the SG $\beta$-disks recover the thermostat property of the
standard disk, namely that the temperature and scale height can
adjust to accomodate (radiate away) the energy input to the
system from viscous dissipation and inward motion.

Second, we note that if the disk matter distribution is clumpy
(e.g., clouds within a low density smoothed out distribution)
then there may be a formal connection between the $\alpha$- and
$\beta$-prescriptions. Since in the $\beta$-formulation the clump
velocities are of order $v_\phi$ shock heating will tend to heat
the low density inter-clump gas until its sound speed $c_{\rm s}
\sim v_\phi$. The inter-clump gas then has a scale height $h \sim
s$, the scale of the clumpy disk, and will hence be roughly a
spherical structure. At this point the $\alpha$- and
$\beta$-prescriptions look formally identical but the scale
height and sound speed now refer to a more or less spherical
background distribution of hot gas in which a disk structure of
cloudy clumps is imbedded.

\subsection{Keplerian Selfgravitating $\beta$-Disks
(KSG)\label{ksgbeta}}

For the particular case of a KSG $\beta$-disk we have from Eq.\
\ref{csquadratsg}

\begin{equation}
\frac{c_{{\rm s}}^2}{h} = \frac{3 G \dot M}{3 \beta} \frac{1}{(G
M_*)^{1/2} s^{1/2}}
\end{equation}
For the SG $\beta$-disk it follows immediately from Eq.\
(\ref{sigmabeta})\ and from mass conservation in the disk that
the radial inflow velocity $v_s$ is given by

\begin{equation}  \label{vssg}
v_s = \frac{\dot M}{2 \pi s \Sigma} = - \beta s^2 \omega^\prime
\end{equation}
For the KSG $\beta$-disk, we then have

\begin{equation}  \label{vsksg}
v_s = \frac{3}{2} \beta s \omega = \frac{3}{2} \beta v_\phi
\end{equation}
Thus at each radius the inward velocity is the same fraction of
the local orbital velocity. From Eq.\ (\ref{vssg}) this, in fact,
holds for any SG $\beta$-disk in which the angular velocity is a
power law function of $s$ with adjustment only to the numerical
factor in Eq.\ (\ref{vsksg}). If $\beta $ satisfies the constraint
(\ref{betalimit}), then the approximation of centrifugal balance
in the radial direction remains well justified.

Under these conditions the dissipation per unit area of a SG
$\beta$-disk is given by

\begin{equation}  \label{dissipation}
D = \frac{\dot M}{4 \pi s} \left(\frac{v_\phi^2}{s}\right) = 2
\sigma T_{{\rm eff}}^4
\end{equation}
where $\sigma$ is the Stefan-Boltzmann constant. For an optically
thick KSG disk this yields the same radial dependence of $T_{{\rm
eff}}$ as for the standard disk, namely

\begin{equation}  \label{teff}
T_{{\rm eff}} = \left(\frac{G \dot M M_*}{8 \pi
\sigma}\right)^{1/4} s^{-3/4}
\end{equation}
This temperature dependence which is identical to that of the
standard model then leads to the well known energy distribution
for an optically thick standard disk of $F_\nu \propto
\nu^{1/3}$. This also implies that---as long as the disks are not
fully selfgravitating---it is hard to distinguish between an
$\alpha$- and a $\beta$-disk model observationally.

\subsection{Fully Selfgravitating $\beta$-Disks
(FSG)\label{fsgbeta}}

We turn now to the case of the fully selfgravitating (FSG)
$\beta$-disk, in which the disk mass is sufficiently great that
it dominates the gravitational terms in the hydrostatic support
equation in both the radial and vertical directions. While there
are many potential solutions for the FSG disk structure, one is
well known in both mathematical and observational terms, namely
the constant velocity ($v_\phi = {\rm const.}$) disk. Within such
a disk structure we have simultaneous solutions to the equation
of radial hydrostatic equilibrium and Poisson's equation of the
form

\begin{equation}  \label{vphisigma}
v_\phi = s \omega = v_0\hspace{4mm}{\rm and}\hspace{4mm} \Sigma
\propto \Sigma_0 \left(\frac{s}{s_0}\right)^{-1}.
\end{equation}
\cite{Too63,Mes63}. For the FSG disk, Eq.\ (\ref{sigmabeta}) then
leads to

\begin{equation}  \label{sigmafsg}
\Sigma = \frac{\dot M}{2 \pi \beta s v_0}
\end{equation}
which has the same radial dependence as the structural solution
shown in Eq.\ (\ref{vphisigma}). Thus Eq.\ (\ref{sigmafsg}) may be
viewed as giving the rate of mass flow through the disk for a FSG
$\beta$-disk with constant rotational velocity $v_0$. Finally,
the equation of continuity provides a constraint if the structure
is to maintain a basically steady state structure. For the
$v_\phi = v_0 = $ const. disk, this yields

\begin{equation}
s\frac{\partial \Sigma}{\partial t} = \frac{\partial}{\partial
s}(s v_s \Sigma) = \frac{\partial}{\partial s}(\beta v_\phi s
\Sigma) = 0
\end{equation}
Thus the constant velocity disk represents a steady state
solution in regions sufficiently far from the inner and outer
boundaries of the $\beta$-disk.

It is then possible, in the spirit of the discussion of Eqs.\
(\ref {dissipation}) and (\ref{teff}), to calculate the energy
dissipation rate per unit area $D$ for the constant velocity
$\beta$-disk. We then find

\begin{equation}
D = 2 \sigma T_{{\rm eff}}^4 = \frac{\dot M v_0^2}{4 \pi s^2}
\end{equation}
so that

\begin{equation}
\label{eqteff}
T_{{\rm eff}} = \left(\frac{\dot M v_0^2}{8 \pi
\sigma} \right)^{1/4} s^{-1/2}
\end{equation}
The flux density, $F_\nu$, emitted by an optically thick,
constant velocity $\beta$-disk it then given by

\begin{equation}
F_\nu \propto \nu^{-1}.
\end{equation}
In reality, a sufficiently massive disk may be expected to have
an inner Keplerian (standard) zone, a Keplerian selfgravitating
zone (KSG), and a fully selfgravitating zo\-ne (FSG). We should
therefore expect a smooth transition in the spectral energy
distribution from the $\nu^{1/3}$ spectrum of the inner two zones
to the $\nu^{-1}$ spectrum arising at longer wavelengths from the
FSG zone. The transition frequency $\nu_{\rm trans}$ may be
derived by solving Eqs.\ (\ref{teff}) and (\ref{eqteff}) for $s$,
determining a value of transition temperature $T_{\rm trans}$ and
setting

\begin{eqnarray}
\nu_{\rm trans} & = & \frac{k T_{\rm trans}}{h} = \frac{k
\dot{M}^{1/4} v_0^{3/2}}{h \left( 8 \pi \sigma\right)^{1/4}
\left( G M\right)^{1/2}} = \nonumber\\ &=&
\left(\frac{k}{h}\right) \left(\frac{\dot{M}}{8 \pi
\sigma}\right)^{1/4} \left(\frac{v_0^3}{GM}\right)^{1/2}.
\end{eqnarray}
One could turn this argument around and argue that, if no other
components contribute to the spectrum, the flatness of the $\nu
F_\nu$ distribution is a measure for the importance of
selfgravity and thus for the relative mass of the accretion disk
as compared to the central accreting object. This, of course,
applies only to the optically thick case which may not arise
frequently in strongly clumped disks.

\subsection{Time Scales\label{secttimescales}}

The evolution of accretion disks can be described by a set of
time scales. For our purposes, the dynamical and the viscous time
scale are of particular interest.

The dynamical time scale $\tau_{{\rm dyn}}$ is given by

\begin{equation}
\tau_{{\rm dyn}} = \frac{1}{\omega}.
\end{equation}
While this formulation applies to all cases, selfgravitating or
not, it is only in the non-SG and in the KSG cases that $\omega$
is given by the mass of the central accretor and by the radius.
In the FSG case, $\omega$ is determined by solving Poisson's
equation.

The time scale of viscous evolution $\tau_{{\rm visc}}$ is given
by

\begin{equation}
\label{equtauvisc}
\tau_{{\rm visc}} = \frac{s^2}{\nu}
\end{equation}
In the standard non-SG and geometrically thin ($h \ll s$) case
($\alpha$ disks), this leads to

\begin{equation}  \label{tauviscnsg}
\tau_{{\rm visc}}^{{\rm non-SG}} = \left(\frac{s}{h}\right)^2
\frac{\tau_{{\rm dyn}}}{\alpha} \gg \frac{\tau_{{\rm
dyn}}}{\alpha}.
\end{equation}
In KSG and FSG disks ($\beta$-disks), $\tau_{{\rm visc}}$ is
given by

\begin{equation}  \label{tauviscsg}
\tau_{{\rm visc}}^{{\rm KSG}} = \tau_{{\rm visc}}^{{\rm FSG}} =
\tau_{{\rm visc}}^{{\rm SG}} = \frac{\tau_{{\rm dyn}}}{\beta}
\end{equation}
With $\alpha < 1$ and $\beta \ll 1$ (Eq.\ \ref{betalimit}) under
all circumstances $\tau_{{\rm visc}} \gg \tau_{{\rm dyn}}$. In
the SG cases the ratio between the two time scales decouples from
the disk structure. In all cases the models are self-consistent
in assuming basic hydrostatic equilibrium in the vertical
direction.

\section{Possible Applications\label{examples}}

\subsection{Protoplanetary Accretion Disks}

T Tauri stars have infrared spectral energy distributions $\nu
F_\nu$ which can be approximated in many cases by power laws $\nu
F_\nu \propto \nu^n$ with a spectral index $n$ in the range $\sim
0 \dots 1.3$. Assuming this spectral behaviour to be due to
radiation from an optically thick disk, it translates into a
radial temperature distribution $T_{{\rm eff}} \propto s^{-q}$
with $n = 4 - 2/q$.

An optically thick non-selfgravitating accretion disk which
radiates energy that is liberated through viscous dissipation,
i.e., an {\it active\/} accretion disk shows a spectral
distribution with $q = 3/4$ or $n = 4/3$. This immediately
excludes optically thick non-selfgravitating standard accretion
disks as the major contributor to T Tau spectra.

\citeasnoun{ALS88} were the first to discuss the possibility of a
non-standard radial temperature distribution with $q \not = 3/4$.
Using $q$ as a free parameter, they find that for flat spectrum
sources, their best fits require disk masses that are no longer
very small compared to the masses of the accreting stars. They
already mention the possibility that the flatness of the spectrum
and selfgravity of the disk may be related. On the other hand, at
that time this indirect argument was the only evidence for large
disk masses. \citeasnoun{BSC90} in their survey of circumstellar
disks around young stellar objects also find preferentially disk
spectra that are considerably flatter than predicted by the
standard optically thick disk models. For more than half of their
objects they derive disk masses that correspond to the KSG and
FSG cases. On the other hand, \citeasnoun{Nat93} proposed that
flat disk spectra are the consequence of dusty envelopes
engulfing a star with a standard disk around it. Recently,
\citeasnoun{CGo97} have investigated in detail
non-selfgravitating {\it passive\/} accretion disks, i.e., disks
that are heated by radiation from the star and re-radiate this
energy. Depending on the details of the flaring of the disk, this
can lead to considerably flatter spectra than expected from
active disks.

However, in the meantime, high resolution direct observations of
protostellar disks yield independent strong evidence for
comparatively large disk masses. \citeasnoun{LCH94}, for
instance, find a lower limit for the disk masses in HL Tau---one
of the sources in Adams, Lada \& Shu's sample of flat spectrum T
Tauri stars---of $\sim 0.02\,{\rm M}_\odot$.

We suggest that the flatness of the spectrum actually reflects
the mass of the disk, i.e., the importance of selfgravity. For
disk masses considerably smaller than $\sim 1/30M_{*}$, the
standard accretion disk models apply. For disks whose masses are
larger but still small compared to $M_{*}$ the spectral behaviour
is not altered significantly, but the disk structure and the time
scale of disk evolution ($\tau _{{\rm visc}}$, see Eqs.\
\ref{tauviscnsg}\ and \ref{tauviscsg}) change. For even more
massive disks, we expect a clear trend towards flatter spectra
that approach an almost constant $\nu F_\nu $ distribution if
selfgravity in the disk becomes important in the radial as well
as in the vertical direction.

\subsection{Galactic Disks}

The relevance of viscosity in the evolution of galactic disks has
been the subject of discussion since \citeasnoun[1951]{vWe43} and
\citeasnoun{Lue52} first raised the issue nearly fifty years ago.
They noted then that, with an eddy viscosity formulation (a
$\beta$-disk), the time scale for evolution of typical galactic
disks was comparable to the age of the universe and suggested
that this might account for the difference between spiral and
elliptical galaxies.

With the subsequent realization that galactic disks moved
primarily under the influence of extended massive halos, interest
in FSG disks waned. However, as noted above, it is possible for a
massive disk to exist and evolve under the influence of
viscosity, while embedded in such a halo gravitational field.
Indeed, in the event that such a structure forms, it must evolve
under viscous dissipation and can achieve a quasi-steady state
with essentially the same mass and energy dissipation
distribution as for the FSG constant velocity disk. We refer to
this case as an Embedded Self-Gravitating (ESG) disk.

The time scale for viscous evolution $\tau_{\rm visc}$ as given in
Sect.\ \ref{secttimescales}\ suggests a means of differentiating
between the $\alpha$- and $\beta$-formulations for this case. For
a normal spiral galaxy with a suggested mean temperature in the
gaseous disk of around $10^4\,$K and a scale height of around
300\,pc, we obtain

\begin{eqnarray}
\tau_{\rm visc}^{(\alpha)} & \sim & 10^4 \tau_{\rm dyn} \sim
3\,10^{11}\,{\rm
yr}\nonumber\\
\tau_{\rm visc}^{(\beta)} & \sim & 10^2 - 10^3 \tau_{\rm dyn}
\sim 3\,10^9 - 3\,10^{10}\,{\rm yr}
\end{eqnarray}
Thus, with these parameters, little evolution would take place in
a Hubble time on the $\alpha$-hypothesis but significant
evolution is predicted on the $\beta$-hypothesis. This problem of
the viscous time scale in a selfgravi\-ta\-ting $\alpha$ accretion
disk was also noted by \citeasnoun[1989]{SBe87}.
\citeasnoun{SFB89} proposed non-axisymmetric disturbances (``bars
within bars") as an alternative way of transporting angular
momentum in the radial direction within a sufficiently short time
scale.

In terms of inflow velocities the $\beta$-ansatz suggests values
in the range $0.3 - 3\,$km\,s$^{-1}$ which would be exceedingly
hard to measure directly. The $\alpha$-ansatz suggests still
lower values. On the other hand, it may be possible to provide
limits on the viscosity through other observational constraints.
For example, the build up of the 3\,kpc molecular ring in our own
galaxy can be interpreted as due to viscosity driven inflow in
the constant velocity part of the galactic disk which ceases (or
at least slows down) in the constant angular velocity inner
regions \cite{Ick79,DBi90}. Similarly, several authors have
suggested that the radial abundance gradients observed in our own
and other disk galaxies may be due to radial motion and diffusive
mixing associated with the turbulence generating the eddy
viscosity \cite{LFa85,SLY90,Koe94,EGr95,TYN95}. According to
these authors, radial inflows of around 1\,km\,s$^{-1}$ at the
galactic location of the Sun are required for optimum fits to the
abundance gradient data within the context of the viscous disk
hypothesis. Such inflow velocities are consistent with the
$\beta$-ansatz but could, of course, be generated also by other
means (e.g effects of bars, magnetic fields).

\subsection{Ultraluminous galaxies}
Recent high resolution imaging of ultraluminous galaxies in the
near infrared and mm wavelengths bands shows dense gas and dust
accretion disks in their galactic nuclei. The two nuclei in the
merger galaxy Arp 220, for instance, have masses of the order of
several $10^9\,{\rm M}_\odot$ within radii of $\lesssim 100\,$pc
\cite{Sco99}. Similar properties, albeit less well resolved than
in Arp 220, seem to be typical for this class of galaxies
\cite{SDR97,DSo98}. Most, if not all, ultraluminous galaxies seem
to be merging galaxies \cite{SMi96}. In Arp 220, these gas masses
are the major contributor to the dynamical mass in the two nuclei
\cite{Sco99}, i.e., these nuclear disks are selfgravitating. Most
likely this is true for the nuclear disks in other ultraluminous
galaxies as well. The merger process is presumably responsible
for transporting large amounts of material into the central few
hundred parsecs, thus filling a mass reservoir which is then
available for subsequent disk accretion to the very center.

Within the framework of $\beta$-disks, one finds that the viscous
accretion time scale $\tau_{\rm visc}$ increases towards larger
radii as long as the surface density $\Sigma$ in the disk
increases with radius $s$ not steeper than $\Sigma \propto s$,
which is most likely fulfilled. Then the viscous time scale at
the disk's outer edge is an upper limit to its evolution time
scale. For Arp 220 (disk mass $\sim 2\,10^9\,{\rm M}_\odot$;
outer radius $\sim 100\,$pc) one finds a time scale of $\sim
10^{5.5}\,{\rm yr}/\beta \sim 3 \dots 30\,10^7$ years for $\beta =
10^{-2 \dots -3}$, which, in turn yields accretion rates $\dot M
\sim 10^{1 \dots 2}\,{\rm M}_\odot\,{\rm yr}^{-1}$. Such rates
lead to accretion luminosities $L_{\rm accr} = \eta \dot M c^2$
up to $\sim \eta 10^{47.5 \dots 48.5}$\,erg\,s$^{-1}$, where
$\eta$ ($\sim 0.1$) is the conversion efficiency of gravitational
energy into radiation and $c$ is the speed of light. Such
luminosities are large enough to power even the strongest AGN and
the time scales are very much shorter than the Hubble time.

Assuming that these rates can be maintained during a sizeable
fraction of $\tau_{\rm visc}$, a significant fraction of the
disk's original gas mass could be accreted to much smaller radii,
presumably to a black hole in the very center (some will be lost
to star formation or winds). In this process the black hole gains
a considerable amount of mass within a relatively short time
scale. One may speculate that this is actually the process that
produces the most massive black holes in the young universe. By
contrast, galaxies that do not undergo mergers presumably have no
way of rapidly collecting such large masses of gas within
$10^2$\,pc. As a consequence, these disks are less likely to be
selfgravitating and thus are likely to have longer $\tau_{\rm
visc}$. The nuclei of such galaxies will accrete much smaller
amounts of material over longer time scale, resulting in lower
mass central black holes \cite[b]{Dus88a}. An example may be our
own Galactic Center.

\section{Summary}
We propose a viscosity prescription based on the assumption that
the effective Reynolds number of the turbulence does not fall
below the critical Reynolds number. In this parametrization the
viscosity is proportional to the azimuthal velocity and the radius
($\beta$-disks). This prescription yields physically consistent
models of both Keplerian and fully selfgravitating accretion
disks. Moreover, for the case of thin disks with sufficiently
small mass, we recover the $\alpha$-disk solution as a limiting
case.

Such $\beta$-disk models may be relevant to protoplanetary
accretion disks as well as to galactic and galactic center disks.
In the case of protoplanetary disks they yield spectra that are
considerably flatter than those due to non-selfgravitating disks,
in better agreement with observed spectra of these objects. In
galactic disks, they result in viscous evolution on time scales
shorter than the Hubble time and thus offer a natural explanation
for an inward flow that could account for the observed chemical
abundance gradients. In galactic centers, $\beta$-disks may be
the supply for powering AGN and for forming supermassive black
holes within time scales short compared to the Hubble time.

Finally, $\beta$-disks yield a natural solution to an
inconsistency in the $\alpha$-disk models if the disk's mass is
large enough for selfgravity to play a role. This problem arises
even in Keplerian selfgravitating disks in which only the vertical
structure is dominated by selfgravity while the azimuthal motion
remains Keplerian.

\begin{acknowledgements}
We thank Fulvio Melia (Tucson), Richard Auer and Achim Traut
(Heidelberg), and the referees Jean-Paul Zahn and Jean-Marc Hur\'e
for helpful comments on the manuscript. WJD acknowledges partial
support by the {\it Deutsche Forschungsgemeinschaft DFG\/}
through {\it SFBs 328 (Evolution of Galaxies)\/} and {439
(Galaxies in the Young Universe)\/}.
\end{acknowledgements}

\appendix

\section{Thermodynamic Considerations for KSG $\alpha$-disks}

For a KSG $\alpha$-disk, we have from Eq.\ \ref{eqtconst}\ that

\begin{equation}
T_{\rm c} = 2.42\,{\rm K} \left( \frac{1}{\alpha}
\frac{\dot{M}}{10^{-6}\,{\rm M}_\odot/{\rm yr}}\right)^{2/3}.
\end{equation}
If the disk is optically thick and advection is negligible,
viscous dissipation leads to local effective temperature of

\begin{eqnarray}
T_{\rm eff} & = & \left( \frac{3}{8\pi\sigma} \right) \left( G M
\right)^{1/4}
\dot M^{1/4} s^{-3/4}\nonumber\\
& = & 8.53\,10^3\,{\rm K} \left(\frac{m \dot m}{s_{\rm
A}^3}\right)^{1/4}
\end{eqnarray}
with $m$ the mass of the central star in solar units and $s_{\rm
A}$ the radius in astronomical units.

An essential thermodynamics requirement is that $T_{\rm c} >
T_{\rm eff}$ or that

\begin{equation}
\frac{T_{\rm eff}}{T_{\rm c}} = 3.53\,10^{-2} \frac{m^{1/4} \dot
m^{-5/12}}{s_{\rm A}^{3/4}} \alpha_{-1}^{2/3} < 1
\end{equation}
This condition is satisfied provided that

\begin{equation}
\dot m > \dot m_{\rm T} = 3.27\,10^{-4} \frac{m^{3/5}
\alpha^{8/5}_{-1}}{s_{\rm A}^{9/5}}
\end{equation}
and that the disk is selfgravitating in the vertical direction at
$s_{\rm A}$. The latter condition leads to a second requirement
on $\dot m$.

For a standard Keplerian disk, the mass flow rate is given (Eqs.\
\ref{hydrostaticnsg}, \ref{nunsg}, \ref{equtauvisc}) by

\begin{equation}
\label{appequmdot}
\dot M \approx \frac{M_{\rm d}}{\tau_{\rm
visc}} \approx \frac{M_{\rm d} \nu}{s^2} = \alpha M_{\rm d}
\left(\frac{h}{s}\right)^2 \omega
\end{equation}
with $M_{\rm d}$ the disk's mass. From Eq.\ \ref{eqcondvertsg},
the condition that the disk is non-selfgravitating is $M_{\rm d}
< (h/2s) M_*$ and hence, from Eq.\ \ref{appequmdot}, that

\begin{equation}
\dot M < \frac{\alpha}{2} \left(\frac{h}{s}\right)^3
\left(\frac{G M_*^3}{s^3} \right)^{1/2}
\end{equation}
or

\begin{equation}
\dot m < \dot m_{\rm G} = 3.14\,10^{-1} \alpha_{-1}
\left(\frac{h}{s}\right)^3 \frac{m^{3/2}}{s_{\rm A}^{3/2}}
\end{equation}
A selfconsistent and physically acceptable solution can be
obtained only if $\dot m_{\rm G} > \dot m_{\rm T}$, that is the
disk becomes selfgravitating at values of $\dot m$ which are
sufficiently high that thermodynamic requirements are not
violated. This condition may then be written as

\begin{equation}
\frac{h}{s} > 1.01\,10^{-1} m^{-3/10} s_{\rm A}^{-1/10}
\alpha_{-1}^{1/5}
\end{equation}
Thus {\it thin\/} KSG $\alpha$-disks appear to be inconsistent
with basic thermodynamic requirements if $m\lesssim 1$, $s_{\rm
A} \lesssim 1$. There is no inconsistency if either or both of
these quantities are sufficiently large.

\end{document}